\documentstyle[twocolumn,aps,floats]{revtex}

\begin{document}

\draft \tolerance = 10000

\setcounter{topnumber}{1}
\renewcommand{\topfraction}{0.9}
\renewcommand{\textfraction}{0.1}
\renewcommand{\floatpagefraction}{0.9}

%Fixing abstract in twocolumn mode
\twocolumn[\hsize\textwidth\columnwidth\hsize\csname
@twocolumnfalse\endcsname

\title{The Theory of Gravitation in the Space - Time with
Fractal Dimensions and Modified Lorents Transformations}
\author{L.Ya.Kobelev  \\ Department of  Physics, Urals State University \\
 Av. Lenina, 51, Ekaterinburg 620083, Russia \\  E-mail: leonid.kobelev@usu.ru}
\maketitle

\begin{abstract}
In the space and the time with a  fractional dimensions the
Lorents transformations  fulfill only as a good approach and
become exact only when dimensions are integer. So the principle of
relativity (it is exact when dimensions are integer) may be
treated also as a  good approximation and may  remain valid ( but
modified ) in case of  small fractional corrections to integer
dimensions of time and space. In this paper presented the
gravitation field theory in the fractal time and space (based on
the fractal theory of time and space developed by author early).
In the theory are taken into account the alteration of Lorents
transformations for case including $v=c$ and are described  the
real gravitational fields with spin equal $2$ in the fractal time
defined on the Riemann or Minkowski measure carrier. In the theory
introduced the new "quasi-spin", given four equations for
gravitational fields (with different "quasi spins" and real and
imaginary energies). For integer dimensions the theory coincide
with Einstein GR or Logunov- Mestvirichvili  gravitation theory.
\end{abstract}

 \vspace{1cm}

%Fixing abstract in twocolumn mode
]

\section{Introduction}

The general relativity theory (GR)\cite{ein} is one of the most known and
used the theories of gravitation field. It is elegant, beautiful from
physical point of view and explains all experimental facts (truly, it must
be point out that others gravitational theories explain them too, as
example see \cite{log}). The one of the known lacks of the GR consists in
impossibility to include other physical fields (except gravitation field)
in the frame of GR, i.e. in impossibility to look on other fields as on
the characteristics of metric of Riemann geometry. The equations of GR
based on  assumption that all systems of frame  are equivalent, the
absolute systems of references are absent thou the space and the time in
the presence of gravitational fields are inhomogeneous. In the fractal
theory of time and space ( see \cite{kob1}-\cite{kob14}) all the physical
fields are included in the fractional dimensions of the time and the
space, the time and the space fields are real fields and any system of
frames are an absolute system of frame, the Lorents transformations and
known physical laws fulfilled as not rigorously laws and are only very
good approach for describing characteristics of the time and the space.
For the  case when fractional corrections to integer dimensions are small
all the equations of the fractal theory practically coincide with
equations of known physical theories or may give only non-sufficient
corrections. Here we present the theory for the fields with spin equal two
in the fractal time and space on the base of the equations given in
\cite{kob1},\cite{kob2},  \cite{kob12} ( in the paper \cite{kob12} used
the modified Lorents transformations with corrections given by the fractal
theory of time  for moving with speed of light(\cite {kob3}-\cite{kob4})
for constructing the theories of scalar, vector and spinor field). The
equation of this theory differs from equations of the theories
\cite{kob10},\cite {kob1} because the modified Lorents transformations
give four systems of gravitational equations for four different
gravitational fields (two with real energies and two with imaginary
energies) instead of only one system in  theories \cite{ein}, \cite{log}.

\section{Generalized fractional derivatives}

Following \cite{kob1}-\cite{kob2}, we will consider both time and space as
the initial real material fields existing in the world and generating all
other physical fields by means of their fractional dimensions. For the
gravitational fields equations construction may be used the principle of
minimum fractal dimensions functional as it was made in
\cite{kob1}-\cite{kob2}. The aim of the paper is to include  (in addition
to Lagrangians used in \cite{kob1}, \cite{kob2}) in Lagrangians of fields
the additional members that give corrections to Lorents transformations
\cite{kob3},\cite{kob4}, \cite{kob12} in the domain of velocities $v\simeq
c $. For describing the functions defined on multifractal sets it is
necessary to introduce the generalized fractional derivatives (see \cite
{kob3}, \cite{kob4}, \cite{kob7}). Therefore, we introduce as in the cited
papers the integral functionals (both left-sided and right-sided) which
are suitable to describe the dynamics of functions defined on multifractal
sets of time and space (generalized fractional derivatives (GFD), see
\cite{kob1}-\cite{kob2}, \cite{kob7}) and replace by  GFD the usual
derivatives and integral respect to time and space coordinates  in the
fractional dimensions functional. These functionals GFD are simple and
natural generalization of the Riemann-Liouville fractional derivatives and
integrals:
\begin{equation} \label{1}
D_{+,t}^{d}f(t)=\left( \frac{d}{dt}\right)^{n}\int_{a}^{t}
\frac{f(t^{\prime})dt^{\prime}}{\Gamma
(n-d(t^{\prime}))(t-t^{\prime})^{d(t^{\prime})-n+1}}
\end{equation}
\begin{equation} \label{2}
D_{-,t}^{d}f(t)=(-1)^{n}\left( \frac{d}{dt}\right)
^{n}\int_{t}^{b}\frac{f(t^{\prime})dt^{\prime}}{\Gamma
(n-d(t^{\prime}))(t^{\prime}-t)^{d(t^{\prime})-n+1}}
\end{equation}
where $\Gamma(x)$ is Euler's gamma function, and $a$ and $b$ are some
constants from $[0,\infty)$. In these definitions, as usually, $n=\{d\}+1$
,where $\{d\}$ is the integer part of $d$ if $d\geq 0$ (i.e. $n-1\le d<n$)
and $n=0$ for $d<0$. If $d=const$, the generalized fractional derivatives
(GFD) (\ref{1})-(\ref{2}) coincide with the Riemann - Liouville fractional
derivatives ($d\geq 0$) or fractional integrals ($d<0$). When
$d=n+\varepsilon (t), \varepsilon (t)\rightarrow 0$, GFD can be
represented by means of integer derivatives and integrals. There are
relations between GFD and ordinary derivatives for $d_{\alpha}$ near
integer values. If $d_{\alpha}$ $\rightarrow n$ where $n$ is an integer ,
( for example $d_{\alpha}$=$1+\varepsilon({\bf r}, (t),t)$, $ \alpha={\bf
r},t )$, in that case it is possible represent GFD by approximate
relations (see \cite{kob1},\cite{kob7})
\begin{equation}\label{3}
  D_{+,x_{\alpha}}^{1+\epsilon}f({\bf r} (t),t)=
  \frac{\partial}{\partial{x_{\alpha}}} f({\bf r} (t),t)+
  \frac{\partial}{\partial x_{\alpha}}[{\varepsilon
  ({\bf r} (t),t)f({\bf r}(t),t)]}
\end{equation}
For $n=1$, i.e. $d=1+\varepsilon$, $\left| \varepsilon \right| <<1$ it is
possible to obtain:
\begin{equation} \label{4}
D_{+,t}^{1+\varepsilon }f(t)\approx \frac{\partial}{\partial t}
f(t)+a\frac{\partial}{\partial t}\left[\varepsilon ({\bf
r}(t),t)f(t)\right]
\end{equation}
where $a$ is constant and defined by the choice of the rules of
regularization of integrals (\ref{1})-(\ref{2}) (for more detail see
\cite{kob1},\cite{kob2}, \cite{kob7}).  The selection of the rule of
regularization that gives a real additives for usual derivative in
(\ref{3}) yield $a=0.5$ for $d<1$ and $a=1.077$ for $d>1$ \cite{kob1}. The
functions under integral sign in (\ref{1})-(\ref{2}) we consider as the
generalized functions defined on the set of the finite functions
\cite{gel}. The notions of GFD, similar to (\ref{1})-(\ref{2}),  also
defined and for the space variables ${\mathbf r}$.\\ In the definitions of
GFD (\ref{1})-(\ref{2})  the connections between fractal dimensions of
time $d_{t}({\mathbf r}(t),t)$ and characteristics of physical fields
(say, potentials $\Phi _{i}({\mathbf r}(t),t),\,i=1,2,..)$ or densities of
Lagrangians $L_{i}$) are determined, following \cite{kob1},  by the
relation
\begin{equation} \label{5}
d_{t}({\mathbf r}(t),t)=1+\sum_{i}\beta_{i}L_{i}(\Phi_{i} ({\mathbf
r}(t),t))
\end{equation}
where $L_{i}$ are densities of energy of physical fields, $\beta_{i}$ are
dimensional constants with physical dimension of $[L_{i}]^{-1}$ (it is
worth to choose $\beta _{i}^{\prime}$ in the form $\beta _{i}^{\prime
}=a^{-1}\beta _{i}$ for the sake of independence from regularization
constant). The definition of time as the system of subsets and definition
the FD $d$ (see ( \ref{5})) connects the value of fractional (fractal)
dimension $d_{t}(r(t),t)$  with each time instant $t$. Thus $d_{t}$
depends both on time $t$ and coordinates ${\mathbf r}$. If $d_{t}=1$ (the
absence of physical fields) the set of time has topological dimension
equal to unity. The multifractal model of time allows, as will be shown
early (\cite{kob3}, \cite{kob4}) , to consider the divergence of energy of
masses moving with speed of light in the special relativity theory as the
result of the requirement of rigorous validity of the conservation laws in
the presence of physical fields that is valid only for closed systems. In
our theory there are an approximate fulfillment of conservation laws as in
the fractal theory of time and space the Universe is  treated as an open
system defined on the measure carrier (the closed system is the Universe
together with the measure carrier).
\\ The gravitational equations may be received by using the principle of
minimum to functional of fractal dimensions with dependencies of GFD
(\cite{kob1}, \cite{kob2}) or by replacing the ordinary derivatives in
proper physical equations by GFD (\cite{kob10}) (the  results will be the
same). In this paper we generalize the theory of gravitational fields
\cite{kob1}, \cite{kob2}, \cite{kob10} by including in the equations
received in these papers the results of paper \cite{kob12} that took into
account the modified Lorents transformations.

\section{The based equations for physical fields with spin equal two}

For generalization of the gravitational field theory presented in
\cite{kob1}, \cite{kob2} by means of construction the equations in the
fractal time and space with modified Lorents transformations we write at
first the field equations for   scalar function $\Phi$ of paper
\cite{kob12}
\begin{eqnarray}\label{6}
(\Box^{2}-4a_{0}^{2}\frac{\partial^{4}}{\partial t^{4}})\Phi({\bf r},t)=
E_{0}^{4}\Phi({\bf r},t)
\end{eqnarray}
where $\Box $ is  D'Alamber operator ($\Box =\Delta
-\frac{\partial^{2}}{\partial t^{2}}$,  $\Delta$  is Laplasian), $\Phi$
are functions describing particles or fields. For scalar  $\Phi$ equation
(\ref{6}) describes the scalar field in the space with fractal dimensions
of time that originate the all physical fields ($a_{0}\neq 0)$ .   The
corrections in (\ref{6}) to the usual D'Alamber equation are the result of
modifying the Lorentz transformations. The last is consequences of fractal
nature of time. Now we may write the equations with taken into account
both phenomenon: the influences of multifractal structure of time ( use in
the equations the generalized Riemann-Liouville fractional derivatives
(GFD) instead of ordinary derivatives) and   corrections to equations from
modified Lorents transformations received in \cite{kob12}. In that case
the equation (\ref{6}) take the form
\begin{eqnarray}\label{7}
(D_{-,t}^{d_{t}}D_{+,t}^{d_{t}}&-& \Delta )^{2}\Phi({\bf r},t) = [
E_{0}^{4}+ \nonumber \\
 &+& 4a_{0}^{2}(D_{-,t}^{d_{t}}D_{+,t}^{d_{t}})^{2}]\Phi({\bf r},t)
\end{eqnarray}
where functionals $D_{+,t}^{d}$ and $D_{-,t}^{d}$ defined by (\ref{1}),
(\ref{2}) and (\ref{5}). It is useful to receive from these equations of
the fourth order (\ref{7})  the four equations of the second order. It is
possible if use the Dirac type four-component matrices $\alpha_{i}$  where
$(i= 1,2,3,4):  \alpha_{i}\alpha_{j}+\alpha_{j}\alpha_{i}=\delta_{ij}$.
  Than we have four equations of second order for the fields both  with
real energies (two equations) and with  imaginary energies (two
equations):
\begin{eqnarray}\label{8}
  (D_{-,t}^{d_{t}}D_{+,t}^{d_{t}}&-& \Delta )I\Phi_{i}({\bf r},t) =
  [\alpha_{1}E_{0}^{2}+ \nonumber \\
 &+& 2a_{0}\alpha_{2}(D_{-,t}^{d_{t}}D_{+,t}^{d_{t}})]\Phi_{i}({\bf r},t)
\end{eqnarray}
where $\alpha_{1}$ and $\alpha_{2}$ may be chosen as in \cite{kob12}
\begin{eqnarray}\label{a}\nonumber
  \alpha_{1}= \left(\begin{array}{cccc}
    1& 0 & 0 & 0 \\
    0 & 1 & 0 & 0 \\
    0 & 0 & -1 & 0 \\
    0 & 0 & 0 & -1 \\
        \end{array}\right) ,  \quad
    \alpha_{2}=\left(  \begin{array}{cccc}
      0& 0 & 0 & 1 \\
      0 & 0 & 1 & 0 \\
      0 & 1 & 0 & 0 \\
      1 & 0 & 0 & 0 \\

    \end{array}\right)
\end{eqnarray}
and $I$ is the  4-component unit matrix. This situation is the same as for
vector, spinor or Proca fields described in the  \cite{kob12}.

\section{The selection of the models of gravity }

The generalization of gravitational theory in fractal space on the base of
 equations (\ref{8}) is possible  by using the two models of
describing the gravity fields : \\ \\
 a)  The first model based on Einstein representations the space-time as the
continuous  set of points described by Riemannian geometry. In  that case
in the fractal theory of time and space the measure carrier must be
defined as the Riemannian sets with integer dimensions. On this sets we
construct the fractal sets of time and space with dimensions defined by
Lagrangians densities of energy for all physical fields (see (\ref{5}) ).
On the fractal sets the laws of physics will broken because of the sets of
space and time become the open systems (see statistical theory of open
system in \cite{klim} ) connected with the measure  carrier.  An arbitrary
theories for any physical fields will include (in the fractal space) the
influences of Riemannian geometry because of Riemannian carrier of
measure. So the Riemannian geometry of fields will be not consequences of
characteristics of fields (for example, gravitational field with spin
equal two), but characteristics of measure carrier. In that case we may
describe the gravitational field (if take into account the influences of
all physical fields on behaviour of gravitational field) by using the
principle of minimum of fractal dimensions functional and Euler equations
with generalized fractional derivatives (GFD) as we introduced in the
equations of paper \cite{kob1}, \cite{kob2}, \cite{kob10}. In this model
it is necessary to use the covariant derivatives in the fractal Riemannian
time-space as it was made in \cite{kob1}, \cite{kob10} and improve them by
introducing the four equations for the gravitational field tensor
$\Phi^{\mu\nu}$ and introducing the corrections from modified Lorents
transformations (the introducing of the last corrections was demonstrated
above). In that case the new gravitational equations will describe four
fields: the two gravitational fields with real energies (if use the
analogy with Dirac theory these fields may have different sign of the
gravitational charges) and the  two gravitational fields with imaginary
energies .  All these fields exist  in the Riemann time-space with fractal
dimensions. In the case of integer dimensions of time and space the
received equations (and the theory) coincide with equations of ordinary
GR. In this paper we consider also the another gravity model with more
analogy to Logunov-Mestvirichvili model of gravity \cite{log}.\\ \\ b) The
second model for describing the gravitation fields in the fractal time and
space ( by GFD using ) consists in the selection of other the measure
carrier. The selection the measure carrier is: the measure carrier
selected as the flat four-dimensions pseudo Euclidean Minkowski
time-space. Our fractal Universe in that case defined as the multifractal
sets on the pseudo Euclidean Minkowski time-space ( the model of a measure
carrier selection in the case a) was the model of Riemannian time-space).
Let us select (as a base) the system of reference which coincide for FD
equal to unit with Cartesian system of reference (we remind  that in the
fractal theory of time and space there are only an absolute systems of
reference but if FD of time and space near integer the principle of
equivalence of all reference systems is valid with grate exactness ). The
equations of the gravitation fields in that case will be similar to the
equations of the theory \cite{log} in which all derivatives replaced on
GFD and metric tensor $\gamma^{\mu\nu}$ are contained the functions
(functionals) originated by fractional dimensions (i.e. it must be the
function of $L$ where $L$ is Lagrangians energy densities of gravitation
fields). Beside these corrections must be taken into account the
corrections (the main corrections) from modifying of Lorents
transformations and the presence as result of it of four sorts of the
gravitational fields (two with real and two with imaginary energies). The
differences the theory of gravitation based on above statements  from the
theory \cite{log} in that case are: 1) the real gravitational fields in
the theory originated by fractional dimensions of time, but not postulated
as in \cite{log}; 2) the ordinary derivatives replaced by GFD for taking
into account the FD of time; 3) we took into account also the modification
of Lorents transformations for time with fractional dimensions; 4) the
time-space is fractal and only the measure carrier is Minkowski time-space
with integer dimensions..

\section{The gravitational equations defined on  Minkowski time-space
 measure carrier}

It is convenient to use the designations of the theory \cite{log} for the
equations construction of  gravitation fractal theory in the multifractal
Universe defined on the pseudo Euclidean Minkowski time-space (this space
may have any dimensions but integer). The equations for gravitation field
tensor $\tilde{\Phi}^{\mu\nu} = \sqrt{-\gamma}\cdot{\Phi}^{\mu\nu}$
(${\gamma}=det(\gamma_{\mu\nu})$, $\tilde{t}^{\mu\nu}$=  $\sqrt{-\gamma}
\cdot{t^{\mu\nu}}$, $L$ - is a Lagrangians density of physical fields (see
in details \cite{log}, \cite{kob1})) have  form
\begin{eqnarray}\label{9}
\gamma^{\alpha\beta}D_{-,\alpha}^{d_{i}}D_{+,\beta}^{d_{i}}I\tilde{\Phi}^{\mu
\nu} = \alpha_{1}b^{2}\tilde{\Phi}^{\mu\nu}+
  \lambda{\tilde t}^{\mu \nu }(\gamma^{\mu\nu},\Phi_{A})+\\ \nonumber
 + 2a_{0}\gamma^{44}D_{-,t}^{d_{i}}D_{+,t}^{d_{i}}\alpha_{2}\tilde{\Phi}^{\mu\nu}
\end{eqnarray}
\begin{eqnarray}\label{A}\nonumber
\tilde{t}^{\mu\nu}=-2\frac{\delta L}{\delta \gamma_{\mu\nu}}
\end{eqnarray}

\begin{equation}\label{10}
  D_{\pm,\mu}^{d_i }\tilde{ \Phi}^{\mu \nu} = 0
\end{equation}
In equations  (\ref{9})  the tensors of gravitational fields
$\tilde{\Phi}^{\mu\nu}$  included in the form of column
\begin{eqnarray}\label{B}\nonumber
  \tilde{\Phi}^{\mu\nu}= \left(\begin{array}{c}
  \tilde{\Phi}_{1}^{\mu\nu}\\
  \tilde{\Phi_{2}}^{\mu\nu} \\
  \tilde{\Phi_{3}}^{\mu\nu} \\
  \tilde{\Phi_{4}}^{\mu\nu} \\
  \end{array}\right)
\end{eqnarray}
The metric tensor $\gamma^{\alpha\beta}$ is a function (or functional) of
the tensors of gravitational fields $\tilde{\Phi}^{\mu\nu}$ and defined on
the fractal pseudo Euclidean Minkowski time-space. These dependencies the
$\gamma^{\alpha\beta}$ from$\tilde{\Phi}^{\mu\nu}$ originated by
dependencies the interval $dS^{2}$ from  $\tilde{\Phi}^{\mu\nu}$ because
the last in the fractal Minkowski space has the complicated form (see
\cite{kob1}) and may be expand in powers of $\tilde{\Phi}^{\mu\nu}$. For
expansion $\gamma^{\mu\nu}$ in powers of $\tilde{\Phi}^{\mu\nu}$ obtain
\begin{eqnarray}\label{11}
\gamma^{\mu\nu}(\tilde{\Phi}^{\mu\nu},D_{+,t}^{d_{t}}\tilde{\Phi}^{\mu\nu},..)
= \gamma^{\mu\nu}+ \\ \nonumber  + \sum
A_{\alpha\beta}^{\mu\nu}\tilde{\Phi}^{\alpha\beta} +...
\end{eqnarray}
were $A_{\alpha\beta}^{\mu\nu}$ are coefficients of expansion and depend
at coordinate and time. So if it is possible when for large distances from
center of gravity $r_{0}$ ($r_{0}<<r$)  to limit oneself by two first
members of (\ref{11})  we may write
\begin{eqnarray}\label{d}\nonumber
 \gamma^{\mu\nu}(\tilde{\Phi}^{\alpha\beta})\approx \gamma^{\mu\nu}+ \tilde{\Phi}^{\mu\nu}
\end{eqnarray}
The equation (\ref{10}) describes the boundary conditions for
$\tilde{\Phi}^{\mu\nu}$ on the Universe surface and $b$ is a mass of
graviton and play role of parameter expanding the domain of existence of
GFD.

 \section{Gravitational fields  defined on the Riemann space measure carrier}

As the carrier of a measure is the Riemann space with an integer
dimensions we obtain the determination for covariant derivatives in
Riemann space with fractional dimensions
\begin{equation}\label{12}
D_{ \pm ,\alpha }^{d_{i} } t^{\mu \nu }  = D_{ \pm ,\alpha }^{d_{i} }
t^{\mu \nu }  + \gamma _{\alpha \beta }^{\nu}  t^{\mu \beta } \;\;\;\; i =
t,r
\end{equation}
where $t^{\mu\nu}$ is the energy-momentum tensor and $\gamma^{\mu\nu}$ is
the metric tensor of the Riemann "four-dimension space with fractional
dimensions", $D_{ \pm,\alpha}^{ d_{i}}$ are GFD, $\gamma _{\alpha \beta
}^{\nu}$ are Christoffel symbols
\begin{equation}\label{13}
\gamma _{\alpha \beta }^\nu  = \frac{1} {2}\gamma ^{\nu \sigma}(D_{\pm
,\alpha }^{d_{i}}\gamma _{\beta\sigma } + D_{\pm ,\beta}^{d_{i}}\gamma
_{\alpha \sigma}+D_{\pm ,\sigma }^{d_{i}} \gamma _{\alpha \beta })
\end{equation}
The equations for gravitation field tensor $\tilde{\Phi}^{\mu\nu}$than
read
\begin{eqnarray}\label{14}
\gamma^{\alpha\beta}D_{-,\alpha}^{\nu,d_{i}}D_{+,\beta}^{\nu,d_{i}}I\tilde{\Phi}^{\mu
\nu}= b^{2}\alpha_{1}\tilde{\Phi}^{\mu \nu }+\lambda{\tilde t}^{\mu \nu }
(\gamma^{\mu\nu},\Phi_{A})+\\ \nonumber +
2a_{0}D_{-,+}^{d_{t}}D_{+,t}^{d_{t}}\alpha{2}\tilde{\Phi}^{\mu\nu}
\end{eqnarray}
where $b$ is a constant value that necessary to introduce for using more
broad sets of functions with GFD and it after calculations must be put
zero. The $\Phi^{\mu\nu}$ is a four column matrix. So we have again four
equations for gravitational fields  with real and imaginary energies. The
equation for curvature tensor (with GFD ) have an usual form, but it will
be four equations for different the curvature tensors $R(i), i=1,2,3,4$
and necessary take into account corrections in the covariant derivatives
from fractal nature of space and of  modifying Lorents transformations
\begin{equation}\label{15}
R_{i}^{\mu\nu}-\frac{1}{2}\gamma^{\mu\nu}R_{i}=\frac{{8\pi}}{{\sqrt{-g}}}T_{i}^{\mu\nu}
\end{equation}
\begin{equation}\label{16}
  D_{\pm,\mu}^{d_i }\tilde g_{i}^{\mu \nu} = 0
\end{equation}
The equation (\ref{16}) describes the boundary conditions for $g^{\mu\nu}$
on the Universe surface. For the case of weak fields the generalized
covariant derivatives may be represented as (see \cite{kob1})
\begin{equation}\label{17}
D_{\pm,\alpha}^{d_i}t^{\mu\nu}\approx\quad 'D_{\pm,\alpha}^{d_i}t^{\mu\nu}
+ ''D_{\pm ,\alpha }^{d_i} t^{\mu\nu}
\end{equation}
The $'D_{\pm,\alpha}^{d_i}$ in (\ref{17}) describes the contribution from
FD of time and space, the member  $''D_{\pm,\alpha}^{d_i}$ describes  the
contribution from Riemann space with integer dimensions.\\ \\ Let us see
what differences between very similar equations (\ref{9}) and (\ref{14}).
The equation (\ref{9}) differs from equation (\ref{14}) based on the
Riemannian measure by three aspects:\\ a) the metric tensor
$\gamma^{\mu\nu}$   in (\ref{9})  determined on the Minkowski space with
fractional dimensions;\\ b)  equations differs  by dependencies of metrics
tensor $\gamma^{\mu\nu}$ from $L$ ( because in the (\ref{9}) there are no
dependencies in the $\gamma^{\mu\nu}$ from $L$ originated by the Riemann
metric tensor), there are only dependencies originating by FD;\\ c) the
reason of appearance in equation (\ref{9}) of the dependencies the
$\gamma^{\mu\nu}$ at $L$ lay in the originate it by the only fractal
dimensions of time and space. If FD are integer the (\ref{9}) coincide
with equation of the theory \cite{log}. If FD integer in (\ref{14}) these
equations coincide with equations of GR. For weak fields GFD may be
represented only by FD covariant derivatives (only two members in the
right part of (\ref{11})) and in that case (\ref{9})  may be represented
by metric tensor $g^{\mu\nu}$ of an "effective" Riemann space with integer
dimensions as in \cite{log} (see also \cite{kob1}).  So (\ref{9}) gives
the equations GR too. We pay attention that the corresponding results of
the theory \cite{log} for connections between metric tensor
$\gamma^{\mu\nu}$ of Minkowski space with "effective" metric tensor
$g^{\mu\nu}$ of Riemann space and gravitation tensor (thou they are valid)
are the special case of our theory. In general case the metric tensor of
Minkowski space are complicated function of gravitation field tensor. We
took into account the alterations in equations originated by modified
Lorents transformations in models with both measures (Riemann and
Minkowski).

\section{Conclusion}

In this paper we considered the two models of gravity theories defined on
the multifractal sets of time and space. From our point of view there are
two main approach to theories of gravity: the first is the approach of
Einstein's theory in which the gravitation fields and forces are no exist
and its role play the curvature of the Riemannian time-space originated by
Riemannian geometry. The second approach is the approach of postulating
the real gravitational fields and forces made in the
Logunov-Mestvirichvili theory  \cite{log}. The last theory treats the
gravity as an usual real field in the flat pseudo Euclidean time-space and
it is the very attractive feature of this theory. The results of both
theories coincide on the distances far from the gravitational radius of
centre of gravity. On the distances of order the gravitational radius (if
our Universe is multifractal set of space and time points defined on the
measure carrier) the both theories it seems are not correct. In the
Universe with multifractal dimensions of time and space on these distances
the main role will play the integral characteristics of GFD and all
equations become not differential but integral equations without
containing of any infinity. We presented in these paper the new theory of
gravity: gravity theory in the time and the space with fractional
dimensions. This theory use the idea and results of works \cite{kob1} -
\cite{kob4}, \cite{kob10}- \cite{kob12}  and take into account the
corrections to SR given by the theory of almost inertial systems in the
time with fractional dimensions \cite{kob11}. In other words  this
gravitation theory expand the main results of \cite{kob12} on the
gravitational fields.\\
\\Let us enumerate now the main results the theory presented in this paper:\\
\\1) The theory gives four sorts of different gravitation fields: two
fields with real energy ( these fields differs by the sign of their
energies (the field for gravitons and the field for anti-gravitons) and
two fields with imaginary energies. The situation is the same as for
vector (electromagnetic) and spinor (electron-positron) fields considered
in \cite{kob12}; \\ \\2) The interactions for each of both fields with
real energies with imaginary energies fields are different. This gives the
possibility to introduce the assumption about existence of new
characteristics of gravitational fields ( "quasi-spin" ) for explaining
these facts;\\
\\ 3) The consideration of two models of a measure carriers ( the measure
on Riemannian space  and the measure on Minkowski space) are made. \\ \\
4) The presented theory coincide with GR or the theory
Logunov-Mestvirichvili for case when FD of time and space become
integer.\\ \\5) In the fractal time and space the ordinary derivatives and
integral must be replaced by GFD. So the main idea of this work may be
used for generalization of all
 gravitational theories not  considered here , including quantum
theories of gravitation.\\ \\6)  In this paper adopted  the point of view:
our Universe is multifractal sets of time and space "points" (see details
in the \cite{kob1}- \cite{kob2}). As any multifractal set it  defined on a
measure carrier. Thus the Universe is an open system (statistical theory
of open systems see in the \cite{klim1}, \cite{klim2}) and the all
physical conservation laws (energy, mass and so on) fulfill as the very
good approach. The exact the conservation laws fulfill only for closed
system: the Universe plus the measure carrier. So the correct selection of
measure is a very serious task and it is the task of near future. In the
domains of Universe where the correction to integer time and space
dimensions are very small (in such domain of Universe we  live and such
domains are in distances far away from gravitational radius) the exchange
by energy, mass, momentum and so on between the Universe and the measure
carrier is very small too and it may be neglected. It is necessary
nevertheless remember about continuous exchange by energy between Universe
and the measure carrier ( absorption and emission of energy) in every
place of our Universe in the frame of presented fractal theory.  The
Universe never lost its energy it seems in that case and the far energy
future of Universe is not so sad.\\
\\  (7) In this paper  we considered the characteristics of gravitational
fields (characteristics  electromagnetic and  electron-positron fields
were considered in \cite{kob12}. Naturally the algorithm used in the paper
may be applied to any fields (electro-weak, Lee-Yang, quarks and so on) in
domains where the fractional correction to the dimensions of time are
small. In that case will be true the main results of this paper: every
physical fields must be replaced by four fields with the real and
imaginary energies. So it seems very likely that all physical fields must
have their imaginary twins (if our Universe is fractal). \\ \\
  8) Nobody knows what the time and the space dimensions  has our
Universe. If the dimensions of time and space are fractional the presented
in this paper the theories of gravitational fields will be true ( if at
least one of the selections the measures carrier are valid) and will
describe the reality of our Universe. As was stressed in \cite{kob11} the
one of methods of verification the fractal theory of time and space is to
accelerate the charge particle to speed of light that in the time with
fractional dimensions  is possible (because for spaces with FD of time the
SR  was modified in the narrow domain of velocities near velocity of light
(see \cite {kob3}-\cite{kob4} ) .

\end{document}